\documentclass[fleqn,usenatbib]{mnras}
\usepackage[utf8]{inputenc}

\usepackage{newtxtext,newtxmath}

\usepackage[T1]{fontenc}
\usepackage{ae,aecompl}
\usepackage{graphicx}
\usepackage{amsmath,amsfonts}
\usepackage{microtype}
\usepackage{booktabs}
\usepackage{siunitx}
\usepackage{geometry}
\usepackage{multirow}
\usepackage{natbib}
\usepackage{subcaption}
\usepackage{journalnames}
\usepackage{hyperref}
\usepackage{comment}
\usepackage{lipsum}
\usepackage[normalem]{ulem}

\DeclareRobustCommand{\VAN}[3]{#2}
\let\VANthebibliography\thebibliography
\def\thebibliography{\DeclareRobustCommand{\VAN}[3]{##3}\VANthebibliography}

\newcommand{\hd}{HD\,49798}
\newcommand{\rx}{RX\,J0648.0$-$4418}

\title[A \nicer\ timing of \hd]{
Timing the X-ray pulsating companion of the hot-subdwarf HD\,49798 with \textit{NICER}
}
\author[M. Rigoselli et al.]{
Michela Rigoselli,$^{1}$\thanks{E-mail: michela.rigoselli@inaf.it}
Davide De Grandis,$^{1}$
Sandro Mereghetti,$^{1}$
and Christian~Malacaria$^{2}$
\\
$^{1}$INAF, Istituto di Astrofisica Spaziale e Fisica Cosmica Milano, via A.\ Corti 12, I-20133 Milano, Italy\\
 $^{2}$ International Space Science Institute, Hallerstrasse 6, 3012 Bern, Switzerland\\
}

\date{Accepted XXX. Received YYY; in original form ZZZ}

\pubyear{2023}

\DeclareSIUnit{\erg}{erg}
\DeclareSIUnit{\year}{y}
\DeclareSIUnit{\pc}{pc}

\newcommand{\nicer}{\textit{NICER}}
\newcommand{\xmm}{\textit{XMM-Newton}}
\newcommand{\rosat}{\textit{ROSAT}}
\newcommand{\apx}[1]{^{\rm #1}}
\newcommand{\pdx}[1]{_{\rm #1}}
\newcommand{\msun}{\ensuremath{\rm{M}_\odot}}
\newcommand{\rsun}{\ensuremath{\rm{R}_\odot}}
\newcommand{\fo}{\hphantom{1}}

\begin{document}
\label{firstpage}
\pagerange{\pageref{firstpage}--\pageref{lastpage}}
\maketitle

    \begin{abstract}
     \hd\ is a hot subdwarf of O spectral type in  a 1.55 day orbit with the X-ray source \rx, a compact object with spin period of   $13.2\,$s. We use recent data from the \nicer\ instrument, joined with archival data from \xmm\ and \rosat, to obtain a phase-connected timing solution spanning $\sim$\,30 years. 
     Contrary to previous works, that relied on parameters determined through optical observations, the new timing solution could be derived using only X-ray data.
    We confirm that the compact object is steadily spinning up with $\dot P=-\num{2.28(2)e-15}\,$s\,s$^{-1}$ and obtain a refined measure of the projected semi-major axis of the compact object $a\pdx{X}\sin i=9.60(5)$\,lightsec. This allows us to determine the inclination and masses of the system as $i=84.5(7)\,$deg, $M\pdx{X}=1.220(8)$\,\msun\ and $M\pdx{opt}=1.41(2)\,$\msun.
     We also study possible long term ($\sim$\,year) and orbital variations of the soft X-ray pulsed flux, without finding evidence for variability. In the light of the new findings, we discuss the nature of the compact object, concluding that the possibility of a neutron star in the subsonic propeller regime is unlikely, while accretion of the subdwarf wind onto a massive white dwarf can explain the observed luminosity and spin-up rate for a wind velocity of $\sim$\,800\,km\,s$^{-1}$. 
    \end{abstract}


\section{Introduction}
\hd/\rx\ is a binary system composed of a pulsating X-ray source with mass $\approx1.2\,$\msun, consistent with either a massive white dwarf (WD) or a neutron star (NS) and a hot subdwarf star of O spectral type and mass $\approx1.4\,$\msun\ \citep{2009Sci...325.1222M}. This makes it peculiar, since it is the only known X-ray  binary with a hot subdwarf mass donor, despite evolutionary models predict the existence of similar systems \citep{1985ApJS...58..661I,1994ApJ...431..264I,2005ARep...49..871Y,2014ApJ...794L..28W,2017ApJ...847...78B}.
The orbital period was determined to be $\approx\SI{1.55}{\day}$ 
in early spectroscopic studies in the optical band \citep{1970MNRAS.150..215T}.
The distance of $(521\pm14)\,$pc has been measured via parallax by \textit{Gaia} EDR3 \citep{2021A&A...649A...1G}.

The spin period of the compact object (13.2 s,  \citealt{1997ApJ...474L..53I}) is decreasing at a steady rate of $|\dot{P}| \approx \SI{2e-15}{\second\per\second}$  \citep{2021MNRAS.504..920M}.
Such a  rapid spin-up cannot be  explained with accretion torques, given the low mass transfer rate ongoing in this binary \citep{2016MNRAS.458.3523M}.
A possible solution has been proposed by \citet{2018MNRAS.474.2750P}, who showed that the observed spin-up is fully consistent with that expected from the contraction of a few million years old WD.

The pulsating  X-ray source shows soft thermal emission, well described by a blackbody with temperature $kT\simeq30$\,eV and emission radius $R\simeq40$\,km, and a power-law tail with photon index $\sim$\,1.8 which dominates the emission above  $\sim$\,0.5\,keV. The total X-ray luminosity of $\approx\SI{e32}{\erg\per\second}$ has not been seen to change appreciably in the $\sim$\,30\,years of monitoring \citep{2021MNRAS.504..920M}.

In this work, we report on observations of \rx\ carried out with the Neutron Star
Interior Composition Explorer (\nicer ) experiment \citep{2016SPIE.9905E..1HG} on board the International Space Station (ISS). 
These data provide, for the first time, a coverage of the whole orbit of the system, thus allowing us to derive the orbital parameters based only on X-ray data. 
In Section~\ref{sec:timing} we use the new \nicer\ data together with the previous \xmm\ and \rosat\ observations to update the phase-connected timing solution. In Section~\ref{sec:flux} we study the variation of the pulsed flux as a function of time and orbital phase; we then discuss our findings and the nature of the compact object in Section~\ref{sec:discussion}.

\section{Observations and data reduction}\label{sec:data}

We base our analysis on an observation campaign conducted by \nicer\ in October 2019, December 2020,  and February 2021 (Table~\ref{tab:obslog}).
In each of the three epochs either one or two orbits of the system were covered. The  sampling of  the source was dense, but not continuous owing to the constraints posed by the $\sim$\,90\,min orbit time of the ISS. The observations provided a total exposure of $\sim$\,99\,ks. The event files were processed using the \texttt{nicerl2} routine of the \texttt{HEASoft} 6.31.1 package (\texttt{NICERDAS V010a}), using the recommended parameters for selecting satisfactory good time intervals ($\texttt{ang\_dist}=0.015$, $\texttt{elv}=15$, $\texttt{br\_earth}=30$) and the latest calibration files available.

\begin{table}
\centering \begin{tabular}{ccc}
\toprule
OBSID & Start time (TDB) & Exposure (s) \\ 
\midrule
\multicolumn{3}{c}{\rosat}\\
\midrule
300226N00 & 1992-11-11T18:07:11 & 5$\,$305 \\
\midrule

\multicolumn{3}{c}{\xmm}\\
\midrule
0112450301 & 2002-05-03T11:40:40 & \fo7\,561\\
0112450401 & 2002-05-04T00:02:51 & \fo7\,636\\
0555460201 & 2008-05-10T21:30:36 & 43$\,$821 \\ 
0721050101 & 2013-11-09T19:32:13 & 39$\,$657 \\ 
0740280101 & 2014-10-18T09:24:27 & 29$\,$645 \\ 
0820220101 & 2018-11-08T11:04:28 & 41$\,$725 \\
0841270101 & 2020-02-27T07:26:26 & 46$\,$326 \\
\midrule
\multicolumn{3}{c}{\nicer}\\
\midrule
2200890101 & 2019-10-04T21:45:10 &  \fo\fo\,999  \\ 
2200890102 & 2019-10-05T02:23:28 &  \fo9$\,$026  \\ 
2200890103 & 2019-10-06T01:37:56 &  \fo5$\,$058  \\ 
3561010101 & 2020-12-02T17:35:54 &  \fo4$\,$678  \\ 
3561010103 & 2020-12-04T00:34:43 & 13$\,$568  \\ 
3561010104 & 2020-12-04T23:55:44 & 15$\,$186  \\ 
3561010102 & 2020-12-02T23:47:21 & 16$\,$394  \\
3561010105 & 2021-02-14T16:30:42 &  \fo2$\,$280  \\ 
3561010106 & 2021-02-15T00:06:58 & 17$\,$247  \\ 
3561010107 & 2021-02-16T00:29:15 & 14$\,$892  \\
\bottomrule
\end{tabular}
\caption{Summary of the data used in this work.}\label{tab:obslog}
\end{table}

For the phase-connected timing analysis, we used also archival data obtained with the \rosat\ PSPC and \xmm\ EPIC-pn instruments (Table~\ref{tab:obslog}), which we analysed as described in \citet{2016MNRAS.458.3523M}.

The \nicer\ and \xmm\ arrival times were barycentered with respect to the solar system using the nominal source position (R.A.$= 6^\text{h}48^\text{m} 4^\text{s}\!\!.6$, Dec.$=-\ang{44;18;58}\!\!\!.\,4$) and the DE405 ephemerides. For \rosat\ we used the DE200 ephemerides, as the newer ones are not available in a format  compatible with the \rosat\ analysis routines. We checked that this results in time differences of order $\approx\SI{1}{\milli\second}$ \citep[e.g.][]{2022Univ....8..360D} that, given the long spin period of the source, $P\simeq\SI{13.2}{\second}$, are well within our errors in the determination of the phase.

\section{Results}
\subsection{Timing analysis}\label{sec:timing}

Similarly to our previous works \citep{2016MNRAS.458.3523M, 2021MNRAS.504..920M}, we derived the timing parameters of \rx\ through the method of phase connection.
To this end, we derived the phases corresponding to the maximum of the pulse profile in the ${[0.2-0.55]\,}$keV  energy range and fitted them with a function composed by a polynomial describing the secular evolution of the spin plus a sinusoidal  modulation to account for the orbital motion, 
    \begin{equation}
        \phi(t)=\nu\,\bar t + \frac{1}{2\,}\dot\nu\,{\bar t}\,^2+A\,\sin\left(\frac{2\pi}{P\pdx{orb}}(\bar t-t^*)\right)\label{eq:spindown}
    \end{equation}
where $\bar t=t-t_0$ is the time shifted by a reference instant $t_0$, ${A=a\pdx{ X}\sin i\cdot\nu/c}$ is the projected semi-major axis expressed in terms of the spin phase, and $t^*$ is chosen in a such a way that the mid eclipse time corresponds to the orbital phase $\Phi\pdx{orb}=0.75$. Note that, in the case at hand, the relative variation of $\nu$ is small enough throughout the considered time-span that $A$ can be treated effectively as a constant within its error. The maxima of the pulse profiles were determined by fitting them with a sinusoidal curve, which is a very good approximation of the observed profile in the considered energy band.

We begin the phase connection procedure starting from the \nicer\ data of 2020, dividing them into ISS orbits (excluding those containing the pulsar eclipse). 
The other \nicer\ data were then gradually included in the fit as the errors of the best fit parameters allowed to maintain phase connection. 
Thanks to the complete coverage of several  orbits of the system, the \nicer\ data allowed us to constrain well the orbital parameters in the sinusoidal term of equation~\ref{eq:spindown}. This had not been possible in previous works, where the phase connected timing solution had been obtained fixing $P\pdx{orb}$ to the value derived from the optical data.

\begin{figure*}
    \centering
    \includegraphics[width=\textwidth]{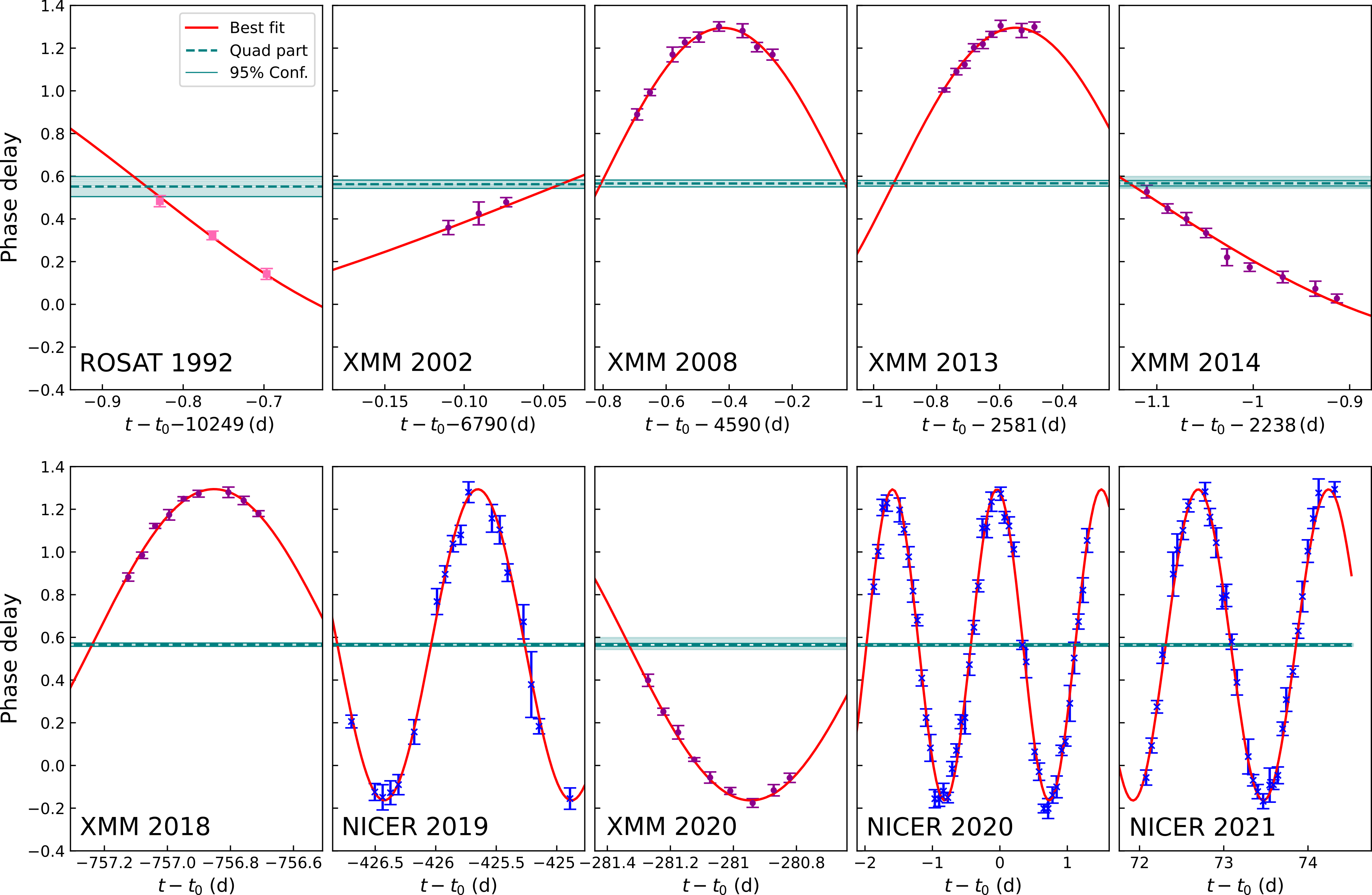}

    \caption{Residuals of the phase connection fit according to equation~\ref{eq:spindown} (red solid curve). The green dashed curve is the quadratic part of the fit (i.e.\ the one describing the spin-up), with the corresponding shaded area indicating the 95\% confidence level uncertainty. 
    }
    \label{fig:connection}
\end{figure*}

\begin{figure}
         \centering
        \includegraphics[width=.5\textwidth]{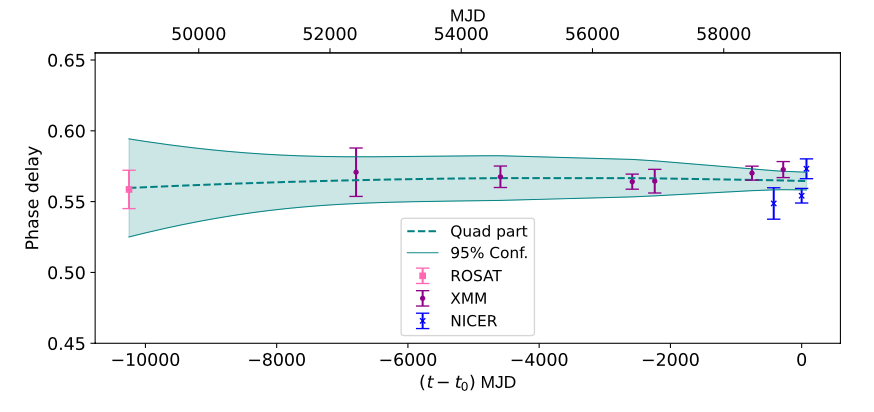}
        \caption{Residuals of the phase connection fit with the sinusoidal term removed. Each point corresponds to the average of the points in one of the panels in Figure~\ref{fig:connection}.}
         \label{fig:sin}
     \end{figure}

We then proceed to connect to this solution   the archival \xmm\ data, that cover a longer timespan. This allows us to constrain the spin frequency derivative $\dot\nu$, which is too small to be measured with the \nicer\ observations only. Therefore, we divided the \xmm\ data in chunks containing a comparable statistics ($\approx500$ counts) and applied the aforementioned fitting procedure to determine the pulse phase. Finally, we added in the same fashion the data from \rosat, thus bringing our baseline to more than $30$ years.
The final fit of all the data is shown in Figures~\ref{fig:connection} and \ref{fig:sin}, and the best fit parameters are given in Table~\ref{tab:timing_sol}.

\begin{table}
    \centering
    \begin{tabular}{ccl}
    \toprule
    Quantity & Value & Unit\\\midrule
        $\chi_\nu^2$& 1.03 for 139 dof&\\\addlinespace[0.4em]
        $t^*$ & 59186.406(1) & MJD (TDB)\\\addlinespace[0.4em] 
        $a\pdx{X}\,\sin i$ & 9.60(5)& lightsec\\\addlinespace[0.4em]
        $P\pdx{orb}$ &1.547666(6)& d \\\addlinespace[0.4em]
        $t_0$ &59187.60882& MJD (TDB)\\\addlinespace[0.4em]
        $\nu$ & 0.07584809567(4) & \si{\hertz}\\\addlinespace[0.4em]
        $\dot{\nu}$&\num{1.31(1)e-17}&{\si{\hertz\per\second}}\\\addlinespace[0.4em]
        $P$&13.184246634(7)&\si{\second}\\\addlinespace[0.4em]
        $\dot{P}$&\num{-2.28(2)e-15}&\si{\second\per\second}\\
        \midrule
        $i$ & $84.5(7)$ & deg\\
        $M\pdx{X\hphantom{x}}$      & $1.220(8) $   & \msun \\
        $M\pdx{opt}$    & $1.41(2)$               & \msun \\
        \bottomrule
    \end{tabular}
    \caption{Best fit parameters of the phase-coherent timing solution (see Figure~\ref{fig:connection}) and updated orbital parameters (see Section~\ref{sec:discussion}).
    }
    \label{tab:timing_sol}
\end{table}

\subsection{Flux stability}\label{sec:flux}

The extensive coverage provided by the \nicer\ observations offers, in principle, the possibility to constrain the variability of the X-ray emission along the orbit, as well as over long timescales. However, given the non-imaging nature of this instrument, this is complicated by two hindrances: the uncertainty in the  background estimate and the presence of another variable source in the field of view. In fact, for this very soft source, most of the flux is at energies close to the lower limit of the \nicer\ band, where the variable and high instrumental background  (in our case even dominant with respect to the source) is difficult to model. Thus, the systematic uncertainties in the background hamper a precise   measurement of the flux of our target.   
Moreover, the X-ray source 3XMM\,J064759.5$-$441941, is located at $\sim$\,1\,arcmin from \hd, well within the field of view of \nicer. This is a main sequence M star with magnitude  $G=12.88$, 
showing chromospheric activity \citep{2020MNRAS.492.4291M}. At energies above $\approx\SI{0.4}{\kilo\eV}$ its flux is higher than that of our target. 
These concerns do not apply if we limit our study to the pulsed flux from \rx .
The total flux can then be inferred under the sole assumption that the pulsed fraction remains constant, as found in all the  previous observations obtained with imaging X-ray instruments.

\begin{figure}
    \centering
    \includegraphics[width=.49\textwidth]{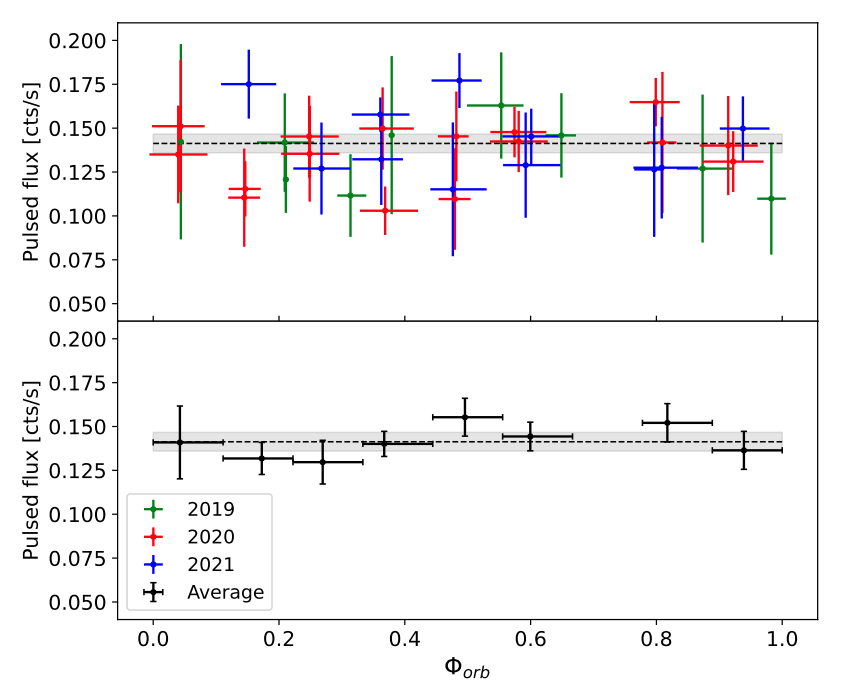}
    \caption{({\it Top panel}): Pulsed flux in the 0.2--0.55 keV band as a function of the orbital phase, excluding the bins containing the eclipse at $\Phi\pdx{orb}=0.75$. Different colours denote the three observations (green: 2019, red: 2020, blue: 2021).
    ({\it Bottom panel}): Same, but averaging all the data in bins of 1/9th of orbital period. The dotted line is the fit with a constant value, with its $1\sigma$ uncertainty shaded in grey.}
    \label{fig:Pflux_phase}
\end{figure}

To this end, we corrected the times of arrival barycentering them with respect to the centre of the \hd/\rx\ system (in addition to the solar system) using the timing solution obtained in the previous section. We then took the counts in the $[0.2-0.55]\,\si{\kilo\eV}$ band and divided them in $9$ bins per orbit, so to have a sufficient statistics in each portion. We folded the resulting chunks of data with the parameters of Table~\ref{tab:timing_sol} and fitted the   pulse profiles with a constant plus a sinusoid. The amplitude of the sine then yields the pulsed fluxes shown in the  upper panel of Figure~\ref{fig:Pflux_phase} as a function of orbital phase.
The lower panel shows the average value obtained by using all the NICER observations folded in 9 orbital bins. 

Aside from the bin containing the eclipse at $\Phi\pdx{orb}=0.75$ (where the value of the pulsed flux is not shown because the pulsations are barely detectable or not visible at all), there is no conclusive evidence of a   variation of the pulsed flux along the orbit, nor between the three time periods at hand. In fact, the fit of all data-points with a constant yields a pulsed flux of $0.141(3)\,$cts~s$^{-1}$ with reduced $\chi^2=0.97$ and, in much the same way, the averaged pulsed fluxes in the three observation periods are compatible with a constant value.

\section{Discussion}\label{sec:discussion}
The phase-connected timing solution derived in Section~\ref{sec:timing} indicates that \rx\ has continued to steadily spin-up also in the most recent years. The new value of $\dot P$ is $\sim$\,5\% higher than the one reported in previous works (the difference is due to a mistake in the conversion of \rosat\ times to TDB); at any rate, this does not affect the conclusion  that such a high spin-up rate cannot be easily explained by accretion torques \citep{2016MNRAS.458.3523M}.
We found values of orbital parameters ($P\pdx{orb}$ and $t^*$) which are fully compatible with the more precise ones determined through optical observations (Schaffenroth et al., in preparation).
In addition, thanks to the complete coverage of all orbital phases provided for the first time by the \nicer\ data, we could refine the value of the projected semiaxis $a\pdx{X}\sin i = (9.60\pm0.05)\,$lightsec. 
This, coupled with $a\pdx{opt} \sin i =( 8.336\pm 0.009)$\,lightsec (Schaffenroth, private communication), implies a mass ratio $q = M\pdx{opt} / M\pdx{X} = 1.153\pm0.006$.

\citet{2013A&A...553A..46M} measured the duration of the X-ray eclipse as $(4311\pm52$)\,s; the inclination of the system can be calculated from this value knowing the radius of \hd, $R\pdx{opt}$. Previous estimates used $R\pdx{opt} = (1.45\pm0.25)\,$\rsun\ \citep{1978A&A....70..653K} and yielded an inclination in the range 79--84 deg. The downward revision of the distance (521 pc instead of 650 pc) and a more recent analysis of the optical/UV spectra, indicate instead $R\pdx{opt} = (1.08\pm0.06)\,$\rsun\ \citep{2019A&A...631A..75K}\footnote{This work relied on an earlier \textit{Gaia} estimate of the distance $d=(508\pm17)\,$pc, resulting in $R\pdx{opt}=(1.05\pm0.06)\,$\rsun.}. This implies an inclination $i = (84.5\pm0.7)\,$deg. 
Finally, using the optical and X-ray mass functions, we can derive updated masses for the two components: $M\pdx{X} = (1.220\pm0.008)$\,\msun\ and $M\pdx{opt} = (1.41\pm0.02)$\,\msun.

Under the assumption that the pulsed fraction of the X-ray emission is not changing over time, our results do not indicate a variation in the flux over the orbit nor between different observations. 
Note that our analysis, limited to the pulsed flux below 0.5 keV, is insensitive to possible variations in the subdominant power-law   component, similar to  that seen between the 2014 and 2020  \xmm\ observations  \citep{2021MNRAS.504..920M}.

\subsection{On the nature of \rx}
Given the orbital parameters of the system, the Roche-lobe of \hd\ has a radius of $\simeq$\,3\,\rsun\ \citep{1983ApJ...268..368E}, significantly larger than that of the hot subdwarf itself. Therefore, accretion cannot proceed through Roche-lobe overflow. 
Nevertheless, given the evidence that \hd\ has a stellar wind with mass loss $\dot M\pdx{w}\simeq\num{2.1e-9}\,$\msun\,yr$^{-1}$ and terminal velocity $v_\infty=1570$\,km\,s$^{-1}$\citep{1981A&A...104..249H, 2019A&A...631A..75K}, it is natural to interpret the observed X-ray flux as the result of wind accretion, whereby only a small fraction, $\epsilon$, of the mass lost by the hot subdwarf is gravitationally captured by the compact object, without the formation of an accretion disk.
In a simple Bondi-Hoyle accretion scenario \citep[e.g.][]{1986bhwd.book.....S} $\epsilon$ can be estimated as 

\begin{equation}\epsilon = \left(\frac{R\pdx{A}}{2a}\right)^2\end{equation}

\noindent
where $a$ is the orbital separation, $R\pdx{A}=2GM\pdx{X}/v^2$ is the accretion radius, and $v = (v\pdx{w}^2 + v\pdx{orb}^2)^{1/2}$  is the relative velocity between the wind and the compact object. For our system, $v\pdx{orb}\simeq 255$\,km\,s$^{-1}$, thus we can take $v\sim v\pdx{w}$.
Since the derived mass does not allow us to discriminate between a NS and a massive WD, in the following we discuss the evidences in favour or against each case. 

\begin{figure*}
    \centering
   \begin{subfigure}{.49\textwidth} \includegraphics[height=8cm]{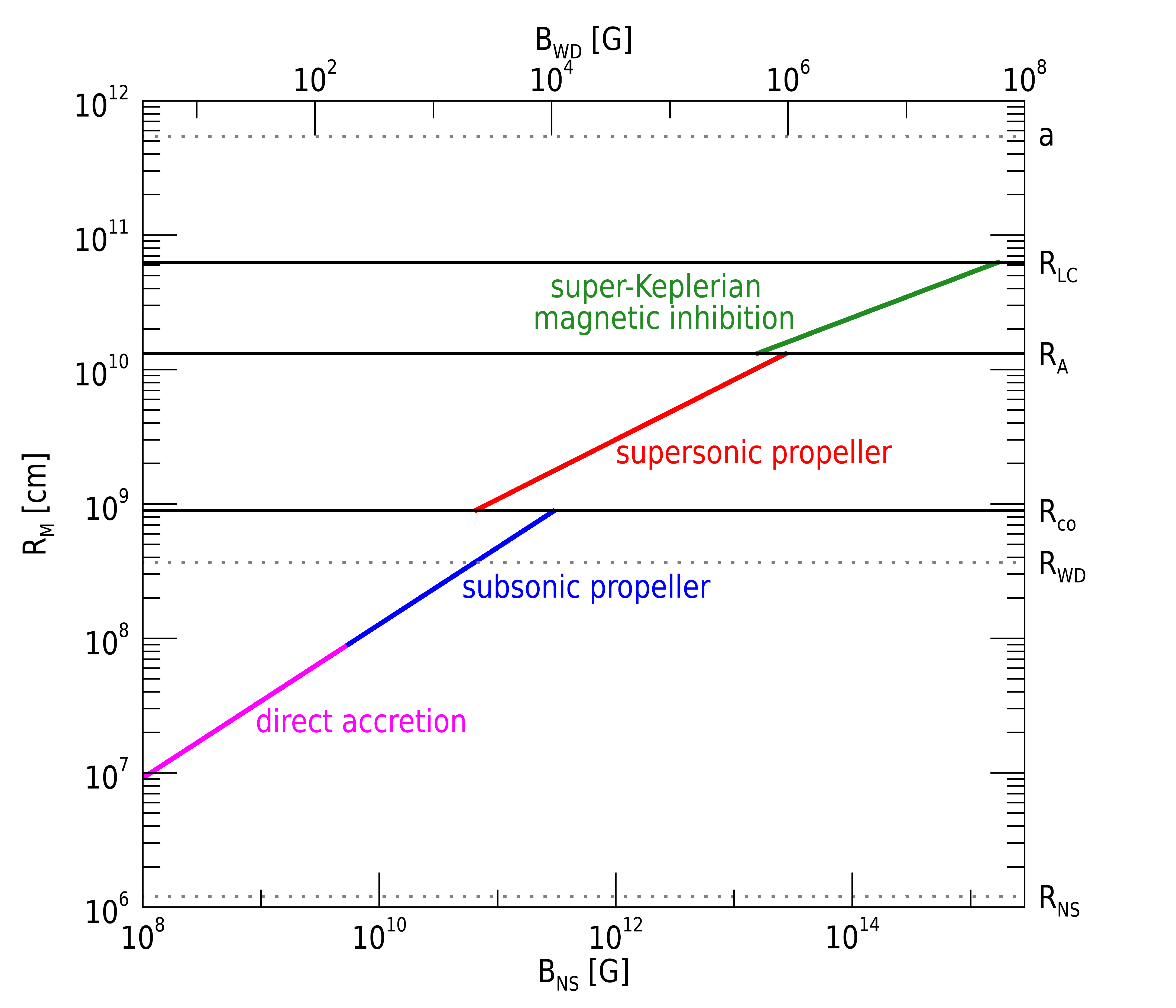}\caption{}\label{fig:RM}\end{subfigure}
   \begin{subfigure}{.49\textwidth} \includegraphics[height=8cm]{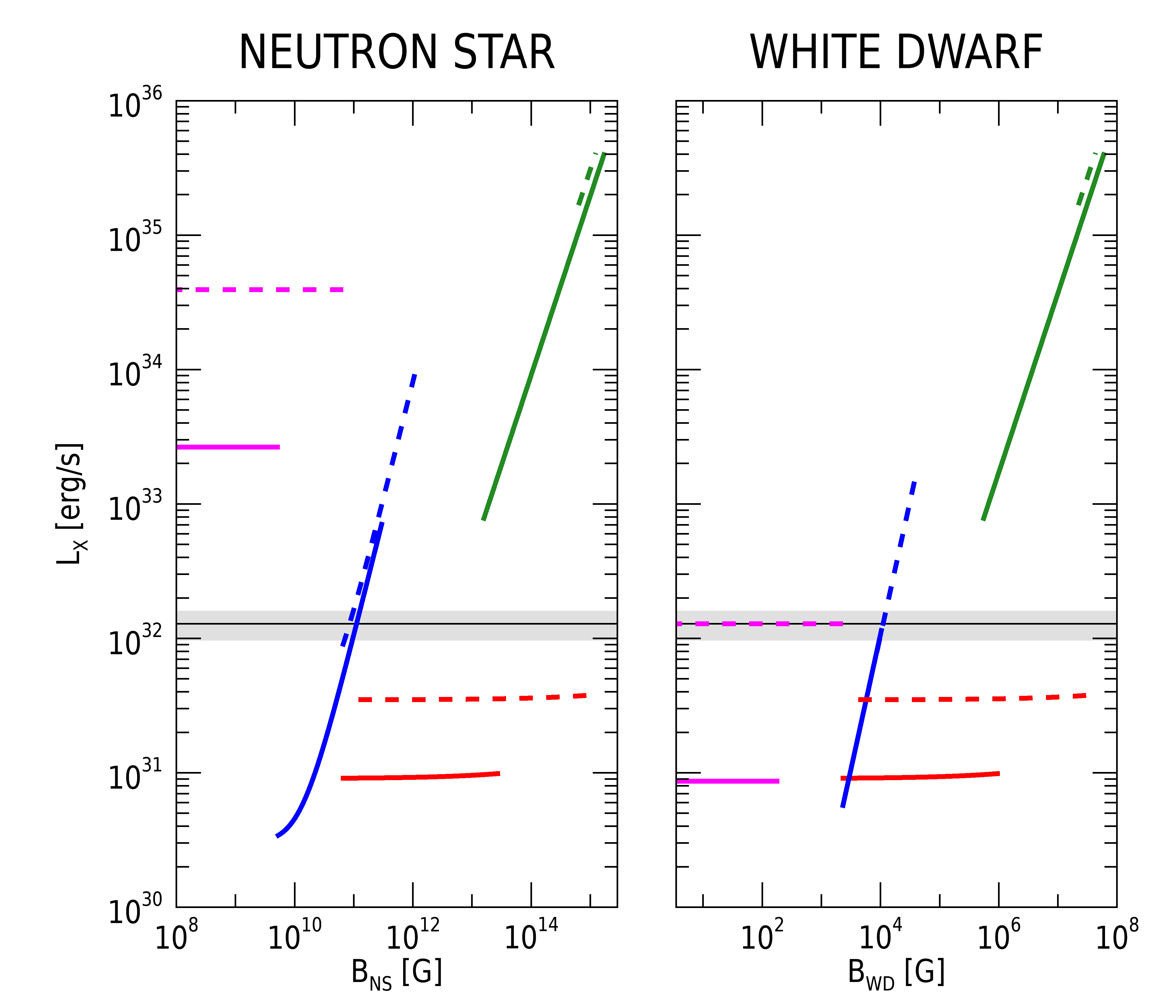}\caption{}\label{fig:LX}\end{subfigure}
    \caption{(\textit{a}):~Magnetospheric radius as a function of magnetic field for a NS (bottom scale) or a WD (top scale). The colors indicate the different regimes: super-Keplerian magnetic inhibition (green), supersonic propeller (red), subsonic propeller (blue), direct accretion (magenta).
    Other characteristic radii of the system are indicated by solid black lines: the light-cylinder radius $R\pdx{LC}=cP/2\pi \simeq 6.3 \times10^{10}$\,cm, the accretion radius $R\pdx{A} = 2GM\pdx{X}/v^2 \simeq 1.3\times10^{10}$\,cm (for $v=1570$\,km\,s$^{-1}$), and the corotation radius $R\pdx{co}= (GM\pdx{X}P^2/4\pi^2)^{1/3} \simeq 9.0 \times10^{8}$\,cm. The grey dotted lines indicate the orbital separation $a\simeq5.4\times 10^{11}$\,cm,  the WD radius $R\pdx{WD}=3580$\,km and the NS radius $R\pdx{NS}=12$\,km.
    (\textit{b}):~The X-ray luminosities expected in the different regimes in the NS (left) and WD (right) cases (colours as in panel (a)) are compared to the observed value $L\pdx{X}= (1.3 \pm 0.3)\times 10^{32}\,$erg s$^{-1}$ (solid black line, error as the grey horizontal band). Solid lines refer to $v=1570$\,km\,s$^{-1}$, while the dashed lines to $v=800$\,km\,s$^{-1}$ (see text).}
   \label{fig:regimes}
\end{figure*}

\subsubsection{Neutron star}
In the aforementioned Bondi-Hoyle scenario, the expected luminosity for accretion onto the surface of a NS with radius $12$ km is 
\begin{equation}
\begin{split}
L\pdx{X}\apx{NS}&=\frac{GM\pdx{X}}{R\pdx{NS}}\,\epsilon \dot{M}\pdx{w} 
 \\\label{eq:LNS}
 &\simeq \num{3e33}\!\left(\frac{\dot M\pdx{w}}{2\times10^{-9}\,\msun/{\rm yr}}\right)\!\left(\frac{v\pdx{w} \vphantom{\dot{M}}}{1500~{\rm km/s}\vphantom{0^9}}\right)^{-4}\,{\rm erg\,s^{-1}},
 \end{split}\end{equation}
which is substantially larger than the observed value, ${L\pdx{X}= (1.3 \pm 0.3)\times 10^{32}\,(d/521\text\,{\rm pc})^2}$\,erg s$^{-1}$ \citep{2016MNRAS.458.3523M}.  

However, the accretion flow that reaches the surface of a NS can be drastically reduced by the effects related to the presence of the rotating magnetosphere. In particular, different regimes are determined by the relative position of four critical radii: the accretion radius $R\pdx{A}$, the corotation radius $R\pdx{co}=(GM\pdx{X}P^2/4\pi^2)^{1/3}$, the light-cylinder radius $R\pdx{LC}=cP/2\pi$ and the magnetospheric radius $R\pdx{M}$ \citep[see, e.g.,][]{1992ans..book.....L}. The value of $R\pdx{M}$ itself acquires a different functional dependence on the parameters of the system in the different   regimes \citep[e.g.][]{1998A&ARv...8..279C,2008ApJ...683.1031B}.
Figure~\ref{fig:RM} displays the behaviour of $R\pdx{M}$ as a function of the magnetic field strength for the parameters of our system, in the accretion regimes that are compatible with them. 
In Figure~\ref{fig:LX} we compare the luminosity expected in the different regimes with the observed value (solid black line, error as the grey horizontal band). 
The magnetic fields have been computed assuming a dipolar configuration; solid lines refer to the terminal wind velocity $v = 1570$\,km~s$^{-1}$, while the dashed lines to $v = 800$\,km~s$^{-1}$.

For NS magnetic fields above $\sim$\,$10^{13}$\,G, the magnetospheric radius is larger than $R\pdx{A}$, resulting in a super-Keplerian magnetic inhibition of the accretion. The wind is thus shocked at $R\pdx{M}$, where the magnetosphere is locally super-Keplerian and supersonic, and the gas kinetic energy is converted into thermal energy. The interaction between the NS magnetic field and matter at $R\pdx{M}$
results in the dissipation of rotational energy and NS spin-down, contrary to what is observed. Furthermore, the corresponding luminosity (green solid line in Figure~\ref{fig:LX}) would be much higher than the observed value.

For lower fields, $5\times10^{9}\,\text{G}\lesssim B\pdx{NS} \lesssim 10^{13}$\,G, $R\pdx{M}$ becomes smaller than $R\pdx{A}$, and the gravitationally captured  matter is expected to form a nearly spherically symmetric and stationary configuration between these two radii \citep{1981MNRAS.196..209D}. If $R\pdx{M} > R\pdx{co}$ (supersonic propeller, red solid lines) a spin-down is still expected, while if $R\pdx{M} < R\pdx{co}$  (subsonic propeller, blue solid lines) a fraction of the inflowing matter can accrete onto the NS, due to the Kelvin-Helmoltz instability\footnote{For the parameters of our system, this mechanism dominates with respect to the accretion deriving from Bohm diffusion \citep{2001A&A...375..944I} for ${B\pdx{NS} \gtrsim \num{2e10}}$\,G.} \citep{1983ApJ...266..175B}.
Using equations 7 and 21 of \citet{2008ApJ...683.1031B} a lower limit on this mass inflow rate can be given as
\begin{equation}
\dot{M}\pdx{KH} \simeq 1.1\times10^{-14}\!\left(\frac{ \epsilon \dot M\pdx{w}}{3.1\times10^{-13}\,\msun/{\rm yr}}\right)^{1/7}\!\left(\frac{B\pdx{NS} \vphantom{\dot{M}}}{10^{11}\,{\rm G}\vphantom{0^9}}\right)^{12/7}\,{\rm \msun/ yr}.
\label{eq:mdotKH}
\end{equation}
With this reduced accretion rate, equation~\ref{eq:LNS} yields a luminosity that is consistent with the observed one for $B\sim10^{11}$\,G.
In this scenario, however, the observed spin-up rate can be barely explained.
In fact, under the most optimistic assumption for the accreted angular momentum, i.e.\ that of matter in a Keplerian orbit at $R\pdx{M}$, the expected spin-up rate would be $\simeq 3\times10^{-17}$ Hz~s$^{-1}$ (for the above $B$ value and a moment of inertia $I\pdx{NS}=\num{e45}\,$g\,cm$^2$).
Nevertheless, the hypothesis of Keplerian rotation at $R\pdx{M}$ is unrealistic, since the angular momentum of the wind matter captured at $R\pdx{A}$, of the order of $\pi R\pdx{A}^2/2P\pdx{orb}$ \citep{2002apa..book.....F},  implies a circularization radius much smaller than $R_M$ for any reasonable wind velocity; therefore, the actual $\dot\nu$ is expected to be significantly lower.

For even lower magnetic fields, $\lesssim 5\times10^9\,$G, the system would be in the regime described by equation \ref{eq:LNS}, which, as discussed above, predicts a luminosity much larger than the observed one (magenta solid lines in Figure~\ref{fig:regimes}).

In conclusion, even considering the interaction between the stellar wind of \hd\ and the magnetic field of \rx\ under different regimes, it is quite difficult to explain the observed X-ray luminosity and spin-up rate. 

At any rate, another element disfavouring the NS hypothesis is that the thermally emitting component has a very large emitting area, radius $\sim$\,40\,km in the case of a blackbody model. This value can only increase by a factor 4--9 when more realistic magnetised atmosphere emission models are applied \citep[see e.g. ][]{ho08}. These dimensions are clearly not compatible with the size of a NS, nor with the idea that the emitting area is located at the edge of the magnetosphere, as this would not account for the large pulsed fraction ($\sim$\,60\% below 0.5\,keV).

\subsubsection{White dwarf}

As the luminosity from equation~\ref{eq:LNS} depends on the radius of the compact object, in case \rx\ is a WD we are faced with the  opposite problem. In fact, for $R\pdx{WD}=\SI{3580}{\kilo\meter}$, computed with the analytic mass-radius relation by \citet{1972ApJ...175..417N}, the luminosity $L\pdx{X}\apx{WD}\simeq\SI{1e31}{\erg\per\second}$ is one order of magnitude below the observed one. 

However, the luminosity has quite a strong dependence on the wind velocity, $L\pdx{X}\propto v\pdx{w}^{-4}$. Therefore, a comparatively small reduction of $v\pdx{w}$ can increase significantly the rate of accreted matter and hence the total luminosity; in our case, a wind velocity of about 800 km~s$^{-1}$ would be required in order to reproduce the observed luminosity (magenta dashed line in the right panel of Figure~\ref{fig:LX}). Indeed, a viable mechanism that could reduce  the wind velocity  is its photo-ionisation by the X-rays emitted from the WD
\citep{2018A&A...620A.150K}.
This mechanism is at work in luminous X-ray sources accreting from the strong winds of massive stars  \citep[e.g.][and references therein]{1991ApJ...379..310S}. It would be interesting to explore this possibility with a self-consistent model, taking into account the peculiar composition of the wind from \hd\ and the way it is influenced by the X-ray emission from \rx.

Even though a mass of $M\pdx{X} = (1.220\pm0.008)$\,\msun\ is quite high within the known population of WDs \citep[e.g.][]{10.1093/mnras/staa2030}, this value, coupled with the fast spin,  is coherent with WD structure theory. In fact, for the period at hand, the mass lower limit imposed by the centrifugal mass-shredding condition is $M\gtrsim 1\,\msun$ \citep[see, e.g., fig.\ 2 of][]{2021A&A...656A..77O}.
The high WD mass has also interesting implications for the future evolution of this system which might lead to a type Ia supernova \citep{2010RAA....10..681W} or to the formation of a NS through accretion-induced collapse \citep{2017ApJ...847...78B}.

\section{Conclusions}
\label{sec:conclusions}

Our new \nicer\ observations of \hd/\rx\ indicate that the compact object in this unique binary has continued its steady spin-up without evidence for long term and/or orbital  variations in the X-ray flux. We also derived new and more accurate values for the system inclination and for the masses of the two components (Table~\ref{tab:timing_sol}), which supersede those first reported in \citet{2009Sci...325.1222M}. 

We thoroughly reexamined the possibility that this system contains a NS, but we could not find a satisfactory scenario able to reproduce at the same time the observed X-ray luminosity, spin-up rate, high pulsed fraction, and large emitting area of the thermal emission. 

The possibility of a WD seems more promising because the spin-up can be accounted for, regardless of the accretion status, by the contraction mechanism proposed by \citet{2018MNRAS.474.2750P}, and there is at least a plausible mechanism, i.e.\ photo-ionisation of the subdwarf wind, that could increase the accretion rate in such a way to be consistent with the observed luminosity. The verification of this scenario deserves a more detailed investigation.

Funnelling of the accreted matter by a low magnetic field (${B\pdx{dip}\lesssim1\,}$kG, and/or the presence of some higher multipoles near the surface) would then account for the formation of a hot spot, thus explaining the high pulsed fraction of the thermal component. The size of the emitting region can extend up to radii of $\sim$\,1000\,km when WD atmosphere models are used in the fit \citep{2021MNRAS.504..920M}.

The high mass of the object $M\pdx{X} = (1.220\pm0.008)$\,\msun\ does not represent an issue in the WD framework: other WDs with similar or even higher mass are known \citep[e.g.][and references therein]{2022AJ....164..131W}. Note however that, contrary to most other cases, the mass of \rx\ has been determined with a dynamical measurement thanks to the presence of X-ray pulsations that make this system equivalent to a double spectroscopic binary. Another remarkable aspect is its short spin period of $13.2$\,s, about twice shorter than that of the next fastest rotating WD \citep{2022MNRAS.509L..31P}.

\citet{2018MNRAS.474.2750P} estimated that there are between 25 and 500 systems similar to \hd/\rx\  in the Galaxy, but these numbers are subject to large uncertainties due to the poorly known properties of the common envelope evolutionary phases. The small distance and high Galactic latitude clearly facilitated the discovery of \hd/\rx , but fainter and more absorbed systems in the Galactic plane might be detected with future X-ray observations.

\section*{Acknowledgements}
We acknowledge financial support from the Italian Ministry for University and Research through grant UnIAM (2017LJ39LM, PI S.~Mereghetti).
We thank V.\ Schaffenroth and J.\ Krti\v{c}ka for providing unpublished results and interesting discussions.

\section*{Data availability}
All the data used in this article are available in public archives.

\bibliographystyle{mnras}
\bibliography{biblio}

\bsp	
\label{lastpage}
\end{document}